\documentclass[10pt,aps,prl,twocolumn,superscriptaddress]{revtex4}

\usepackage[utf8]{inputenc}
\usepackage{hyperref}
\usepackage{graphicx}
\usepackage{amsmath,amssymb}
\usepackage{color}
\usepackage{subfigure}
\usepackage{url}

\graphicspath{{Images/}}

\begin{document}

\title{Singular value demodulation of phase-shifted holograms
}

\author{Fernando Lopes}
\affiliation{
Institut de Physique du Globe de Paris - Pal\'eomagn\'etisme, Universit\'e Paris Diderot, Sorbonne Paris Cit\'e, 1 rue Jussieu, Paris, France
}

\author{Michael Atlan}
\affiliation{
Centre National de la Recherche Scientifique (CNRS) UMR 7587, Institut Langevin. Fondation Pierre-Gilles de Gennes, Institut National de la Sant\'e et de la Recherche M\'edicale (INSERM) U 979, Universit\'e Pierre et Marie Curie (UPMC), Universit\'e Paris 7. \'Ecole Sup\'erieure de Physique et de Chimie Industrielles ESPCI ParisTech - 1 rue Jussieu. 75005 Paris. France\\
}

\date{\today}

\begin{abstract}
We report on phase-shifted holographic interferogram demodulation by singular value decomposition. Numerical processing of optically-acquired interferograms over several modulation periods was performed in two steps : 1- rendering of off-axis complex-valued holograms by Fresnel transformation of the interferograms; 2- eigenvalue spectrum assessment of the lag-covariance matrix of hologram pixels. Experimental results in low-light recording conditions were compared with demodulation by Fourier analysis, in the presence of random phase drifts.
\end{abstract}

\maketitle

Recording optical holograms by time-modulated interferometry was proposed and demonstrated in early studies~\cite{Macovski1969, Aleksoff1969, Dandliker1980}. It pioneered the development of phase-shifting~\cite{Yamaguchi1997} holographic interferometry. Time-modulated phase-shifting interferometry involves recording of a sequence of interferograms over at least one modulation cycle of the intensity. Temporal demodulation consists in a linear combination of a sequence of recorded interferograms, whose coefficients are the ones of a discrete Fourier transform~\cite{Creath1985, Freischlad1990, Schnars2002}. An important feature of holographic interferometry is to allow spatial signal modulation in off-axis recording conditions~\cite{LeithUpatnieks1962}, which permits the discrimination of self-beating and cross-beating interferometric contributions of the object field and the reference field. In this configuration, spurious interferometric contributions can be filtered-off~\cite{Cuche2000}. It was demonstrated that the combination of spatial and temporal signal modulation enables low-light imaging at shot noise levels~\cite{GrossAtlan2007, LesaffreVerrierGross2013}. Important features and applications include optical phase imaging~\cite{CucheMarquet1999, CharriereColomb2006, Pandey2011, CotteToy2013} and detection tunability in the radiofrequency range for optical heterodyne detection~\cite{BrunoLaurentRoyer2014}, frequency-resolved narrowband sensing of scattered optical radiation~\cite{GrossDunn2005, AtlanDesbiolles2010, RuanMatherMorgan2013}. Sensitivity is compatible with photon-counting recording conditions~\cite{YamamotoYamamoto2009, LatorreCarmonaJavidi2013, DemoliSkenderovic2014}. In standard phase-shifting interferometry, a sequence of interferograms with known phase shifts between probed and reference optical waves is recorded. When the modulation is not known, principal component analysis~\cite{Vargas2011} can be used for non-parametric, non-iterative demodulation. It identifies uncorrelated variables in the structure of the data (the principal components) from the analysis of its correlations. Non-parametric approaches were also applied to image formation concepts~\cite{BerteroPike1982}, speckle reduction~\cite{MarengoRodriguezFederico2007, BerniniGalizzi2007, EquisJacquot2009, SuLeeLee2010, LeoPiccolo2014}, and laser speckle contrast analysis of flows~\cite{HumeauHeurtierAbrahamMahe2014}.\\

In this letter, we demonstrate experimentally that the sensitivity of low-light optical signal retrieval in time-modulated holographic interferometry can be improved with respect to Fourier transform demodulation in practical conditions, when random signal fluctuations are present. For that purpose, a data-driven, non-parametric signal processing method is used. It consists of calculating the spectrum of eigenvalues in a singular value decomposition (SVD) of the lag-covariance matrix of recorded temporal fluctuations of the hologram's pixels, after spatial Fresnel transformation. The proposed processing scheme is less affected by random optical phase drifts occurring during the sampling process than signal demodulation by discrete Fourier transformation.\\

In optical heterodyne detection, a field of interest $E(t) = {\cal E}  \exp \left( i \omega_{\rm L} t \right)$, oscillating at the optical frequency $\omega_{\rm L}$, is non-linearly mixed with a local oscillator field $E_{\rm LO}(t) = {\cal E}_{\rm LO} \exp \left( i \Delta \omega t \right) \exp \left( i \omega_{\rm L} t \right) $ that is set at a close-by intermediate frequency $\omega_{\rm L} + \Delta \omega$. In this notation, ${\cal E}$ and ${\cal E}_{\rm LO}$ are complex constants and $i$ is the imaginary unit. The amplitude and phase of the original signal $E(t)$ can be retrieved in the desired outcome oscillating at the difference frequency $\Delta \omega$, which can be set within the receiver's temporal bandwidth. The frequency conversion process is ensured by non-linear detection of the optical field $E$ by the array of square-law sensors of a camera, which respond quadratically with the impinging electric field's magnitude. The squared magnitude of the total field received, $I(t)  =  \left| E(t) + E_{\rm LO}(t) \right|^2$, has cross-terms oscillating at the difference frequency $\Delta \omega$ of the fields $E$ and $E_{\rm LO}$, causing modulation of the optical energy flux $I(t) = \left| {\cal E} \right|^2 + \left| {\cal E}_{\rm LO} \right|^2 + {\cal E} {\cal E}_{\rm LO}^* e^{- i \Delta \omega t} + {\cal E}^* {\cal E}_{\rm LO} e^{ i \Delta \omega t}$, where $^*$ denotes the complex conjugate. This equation describes the temporal fluctuation of the recorded intensity at a given pixel. Holographic image rendering, referred to as spatial demodulation, is then performed onto each recorded interferogram $I$ by a discrete Fresnel transform. This transformation involves the free-space propagation impulse response of the optical field~\cite{Schnars2002, KimYuMann2006}. In off-axis recording configuration, the Fresnel transform separates spatially the four interferometric terms of $I(t)$~\cite{LeithUpatnieks1962, Cuche2000} and forms a reconstructed hologram. In the absence of random fluctuations, the pixels of the off-axis part of the hologram carry the signal
\begin{equation}
X(t) = {\cal E} {\cal E}_{\rm LO}^*\exp(- i \Delta \omega t)
\label{eq_X}
\end{equation}

The purpose of signal demodulation is to retrieve the magnitude of the oscillating signal at the frequency $\Delta \omega$ at each pixel. In ideal conditions, Eq.~\ref{eq_X} holds and signal demodulation by discrete Fourier analysis of the recorded interferograms~\cite{Freischlad1990, Schnars2002} enables shot-noise limited detection in dim light~\cite{GrossAtlan2007}; yet in actual experiments, random amplitude and phase fluctuations limit the detection sensitivity.\\

\begin{figure}[]
\centering
\includegraphics[width = 7 cm]{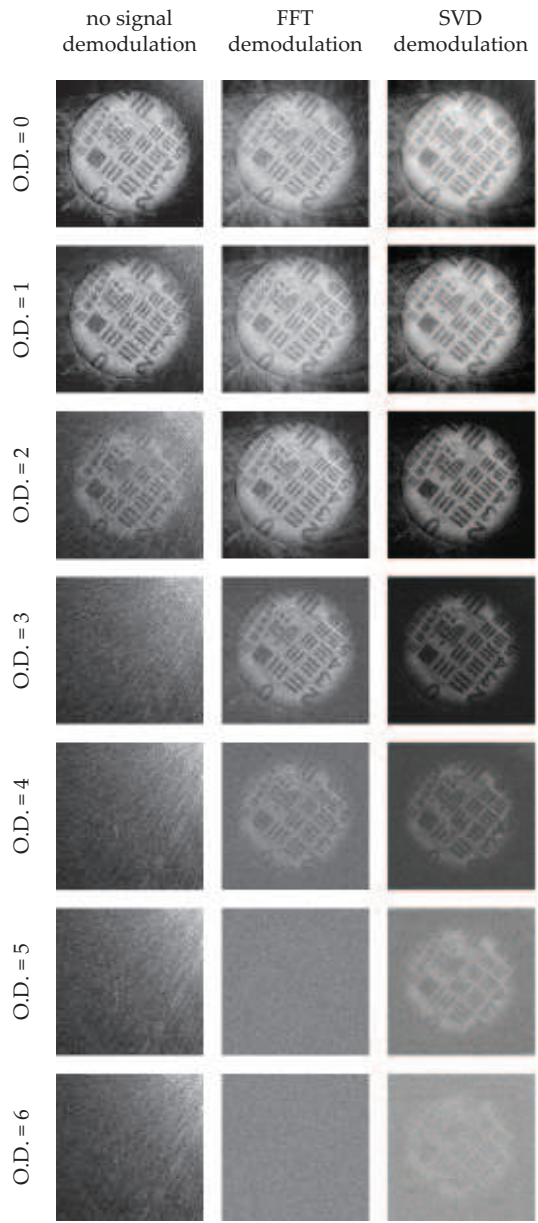}
\caption{Amplitude holographic images from 4-phase modulated interferograms acquired over 128 frames. No time demodulation (first column). Demodulation by Fourier transform (second column). Demodulation by singular value decomposition (third column).}
\label{fig_LowLightImages_RAW_FFT_SVD}
\end{figure} 

We recorded phase-shifted interferograms over several modulation periods in low-light and studied signal retrieval procedures. Optical recording of interferograms was performed with a frequency-shifted, off-axis Mach-Zehnder interferometer~\cite{AtlanGross2007} used for optical heterodyne detection of the object field $E$, backscattered from a resolution target, beating against a separate local oscillator field $E_{\rm LO}$, on a sensor array. The main optical field was provided by a 150 mW, single-mode laser (wavelength $\lambda = 532$ nm, optical frequency $\omega_{\rm L}/(2 \pi) = 5.6 \times 10^{14} \, \rm Hz$, Cobolt Samba-TFB-150). The optical frequency of the local oscillator beam was shifted by a tunable quantity $\Delta \omega$ by two acousto-optic modulators (AA-electronics, MT200-BG9-FIO) driven by phase-locked continuous-wave radiofrequency signals around 200 MHz. The observed reflective resolution target was illuminated over a $\sim 10 \, {\rm mm}$-diameter disk. Interferograms were measured with a Ximea MQ042MG-CM camera ($2048 \times 2048$ pixels CMOSIS CMV4000 sensor array, pixel size $d = 5.5 \, \mu \rm m$, full well charge : 13500 $e^-$ (photo-electrons), conversion gain : $1/G = 0,075 \,{\rm counts}/e^-$, quantum efficiency at 532 nm $\eta \sim 0.5$), running at an externally-triggered frame rate of $\omega_{\rm S} / (2 \pi) = 20 \, \rm Hz$, at 8 bit/pixel quantization. In the reported experiment, a reflective USAF target printed on white paper was illuminated in reflection, and the recorded intensity of the light in the object channel accounted for $11.56 \, {\rm counts} \times G \simeq 154$ photo-electrons per pixel per frame, on average. The average intensity of the light in the LO channel accounted for $63.89 \, {\rm counts} \times G \simeq 852$ photo-electrons per pixel per recorded frame. In these conditions, a sequence of 128 consecutive raw interferograms was recorded for a detuning frequency of the local oscillator of $\Delta \omega/(2\pi) = 5 \, \rm Hz$, creating 4-phase ($\omega_{\rm S}/\Delta\omega = 4$) modulation conditions~\cite{AtlanGross2007}. Twelve more sequences of images were acquired after decreasing the optical power in the object channel by $10^{0.5}$ from one measurement to the next, by combining optical densities (O.D., Thorlabs NE absorptive neutral density filters). The transmitted power to the target scaled as $10^{- \rm O.D.}$, from $10^0$ to $10^{-6}$; hence the average number of recorded photo-electrons per pixel and per frame in the object channel ranged from 154 to $1.54 \times 10^{-4}$. The LO-to-object optical power ratio (heterodyne gain) ranged from  $G_{\rm H} \simeq 5.5 $ to $G_{\rm H} \simeq 5.5 \times 10^6$, because the LO power was kept constant throughout the experiment. The holographic images of these two datasets were then demodulated spatially~\cite{KimYuMann2006}, to yield $661 \times 661$ complex-valued time traces $X(t)$ of 128 elements each, cropped from the $2048\times 2048$ calculation array, over an off-axis spatial region of interest where the image was expected. We then compared two types of signal demodulation procedures to assess the magnitude of the oscillating signal at the frequency $\Delta \omega$ in $X(t)$ at each pixel : Fourier analysis, and singular value analysis of temporal fluctuations.\\

\begin{figure}[]
\centering
\includegraphics[width = 7 cm]{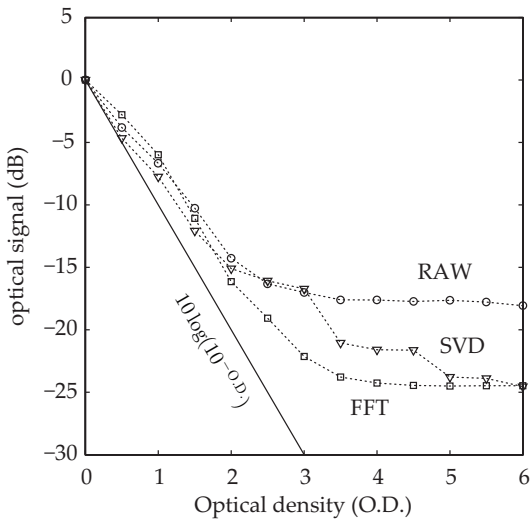}
\caption{Signal magnitude versus optical density (O.D.) in the object channel from a sequence of $N = 128$ frames of 4-phase interferograms. No time demodulation (circles). Demodulation by discrete Fourier transform (squares). Demodulation by singular value decomposition (triangles).}
\label{fig_RAW_FFT_SVD_RadiometryPlots}
\end{figure} 

\begin{figure}[]
\centering
\includegraphics[width = 8 cm]{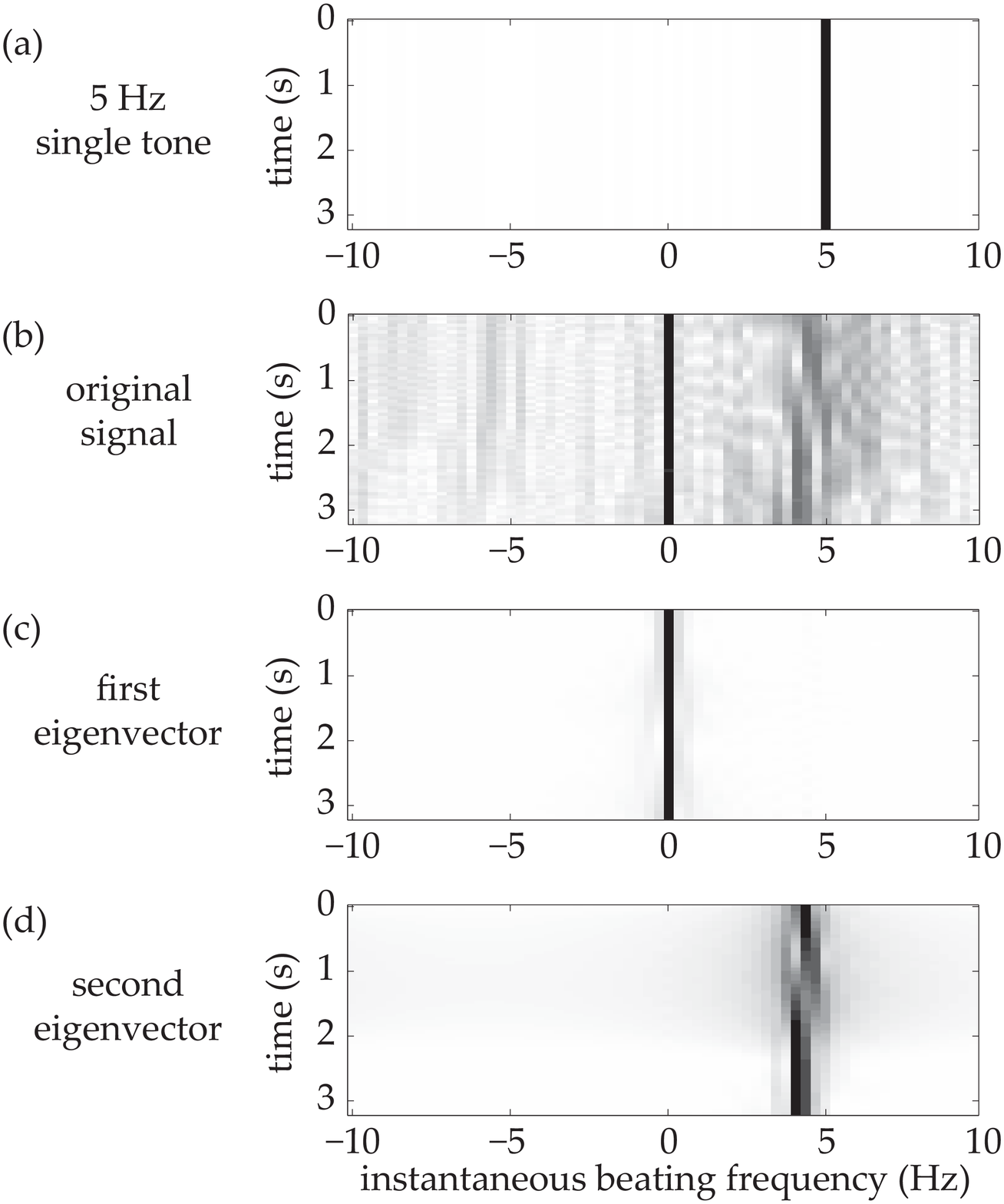}
\caption{Spectrograms of a complex wave oscillating at $5\,\rm Hz$ (a), a hologram's pixel signal $X(t)$ (b), signals reconstructed from the first (c) and second (d) elements of the SVD of $D$.}
\label{fig_Spectrograms}
\end{figure}

First, holographic images were rendered without demodulation. The squared magnitude of the object field was averaged over $N=128$ frame sequences, and reported in the first column of Fig.~\ref{fig_LowLightImages_RAW_FFT_SVD}. The image is not visible for optical power transmission factors lower or equal to $10^{-3}$ (i.e. ${\rm O.D.} = 3$), which corresponds to $19.7$ photo-electrons per pixel recorded on average during 128 frames. Then, the recorded optical fluctuations were projected on the expected oscillation to retrieve the modulated component to improve the detection sensitivity. The modulation depth of the signal was assessed by a fast Fourier transform (FFT) $\tilde{X}$ of the sequence $x_k$ at the modulation frequency $\omega_{\rm S}/4$ (5 Hz), calculated on 128 time points. Results are reported in the second column of Fig.~\ref{fig_LowLightImages_RAW_FFT_SVD}. The image is not visible for a power transmission factor lower or equal to $10^{-5}$ (i.e. ${\rm O.D.} = 5$), which corresponds to $1.97\times10^{-1}$ photo-electron per pixel recorded on average during 128 frames. Finally, we performed singular value analysis of temporal fluctuations. In this approach, we embedded the time series $X=\{x_1, ..., x_N\}$ of the signal at each pixel in a Hankel matrix of dimension $K \times (N-K+1)$, where $K$ is the number of principal components to seek, and $N=128$ is the total number of acquired interferograms. The Hankel matrix $D$, composed of $N-K+1$ lag-shifted copies of truncated values of $X(t)$ has the form
\begin{equation}\label{eq_HankelMatrix}
D = \begin{bmatrix} 
x_1 & x_{2} & \dots  & x_{N-K+1} \\ 
x_{2} & x_3 & \ddots & \vdots \\ 
\vdots & \vdots & \ddots  &  x_{N-1}\\ 
x_{K} & x_{K+1} & \dots  & x_N \\ 
\end{bmatrix}
\end{equation}
We proceeded by diagonalizing the lag-covariance matrix of $X(t)$, $C_X = D D^{\rm t*}$, where $D^{\rm t*}$ is the adjoint matrix (Hermitian conjugate) of $D$. The $K$ eigenvalues $\lambda_k, k = 1,...,K$ of $C_{X}$, numbered by decreasing order of magnitude, were calculated by SVD~\cite{GolubReinsch1970} of $C_X$. The length of the analysis window $K$ was chosen to be an integer number of expected periods of the modulated signal~\cite{VautardGhil1989} $K = p \, \omega_{\rm S}/\Delta\omega$. We used a window length of $p=5$ modulation periods ($K = 20$). Images were rendered by calculating the amplitude of the second eigenvalue at each pixel. They are displayed in Fig.~\ref{fig_LowLightImages_RAW_FFT_SVD} in the third column, as a function of the optical density in the object channel. With this approach, the image is visible down to an optical power transmission factor of $10^{-6}$ (i.e. ${\rm O.D.} = 6$), for which the signal wave intensity accounted for a total of $1.97 \times 10^{-2}$ recorded photo-electron per pixel during 128 frames, on average. The algorithm used for the SVD is based on  the orthogonal triangular decomposition (QR factorization)~\cite{GolubVanLoan2012}. SVD and FFT were performed by the functions \emph{svd()} and \emph{fft()} in Matlab 2012a, run on a CentOS 6 Linux 64-bit operating system. The response curve of singular value demodulation was assessed by plotting the quantity $10\log_{10}(\left<S\right>/\left<S_0\right>)$ (in dB) against the value of the optical density in the object channel (Fig.\ref{fig_RAW_FFT_SVD_RadiometryPlots}). The braces $\left< \,\right>$ indicate spatial averaging of $S$, and $S_0$ within and outside the reconstructed image, respectively. The expected ideal detection behavior is plotted as a continuous black line (-10 dB of optical power per unit of optical density). The demodulation by SVD is qualitative~\cite{BroomheadKing1986}; yet it enables to retrieve images at much lower signal levels than FFT demodulation. The singular value demodulation procedure is illustrated by the spectrograms of four signals, reported in Fig.~\ref{fig_Spectrograms} : (a) the complex wave oscillating at $\Delta\omega /(2\pi)=5\,\rm Hz$, which would be expected in perfect modulation conditions, in the absence of random phase fluctuation (Eq.~\ref{eq_X}); (b) the actual signal $X(t)$ from a pixel at O.D. = 2, which exhibits a DC contribution and jitter - deviation from the modulation frequency; and the signal reconstructed from the first (c) and second (d) eigenvectors and eigenvalues of the SVD of $D$. In this example, the DC term is found in the first component, and the modulated component at 5 Hz undergoing a phase drift is retrieved in the second component of the SVD.\\

In conclusion, we used heuristically a singular value decomposition procedure to retrieve time-modulated optical signals in holographic interferometry. In the reported experiment, over long acquisition times, singular value demodulation of phase-shifted interferograms enabled qualitative image visualization at much lower signal levels than demodulation by discrete Fourier transformation. This method will be investigated for signal retrieval in strong phase fluctuations conditions, phase microscopy, and singular spectrum analysis of pseudoperiodic modulations.\\

This work was supported by Agence Nationale de la Recherche (ANR-09-JCJC-0113, ANR-11-EMMA-046), Fondation Pierre-Gilles de Gennes (FPGG014), r\'egion Ile-de-France (C’Nano, AIMA), the Investments for the Future program (LabEx WIFI: ANR-10-LABX-24, ANR-10-IDEX-0001-02 PSL*), and European Research Council (ERC Synergy HELMHOLTZ).

\end{document}